\documentstyle{memsait}
\newcommand{\tab}[1]{Tab.\,\ref{#1}}
\newcommand{\abb}[1]{Fig.\,\ref{#1}}
\newcommand{\mzams}{M_{\rm ZAMS}}
\newcommand{\etal}{et al.\ }
\newcommand{\msun}{\, {\rm M}_\odot}
\newcommand{\cdr}{^{13}\textrm{C}}
\newcommand{\nvi}{^{14}\textrm{N}}
\input epsf.sty

\begin{opening}
\title{TP-AGB EVOLUTION WITH OVERSHOOT FOR LOW-MASS STARS AS A
FUNCTION OF METALLICITY} 
\author{FALK HERWIG$^1$, THOMAS BL\"OCKER$^2$, THOMAS DRIEBE$^2$}
\institute{$^1$Universit\"at Potsdam, Institut f\"ur Physik,
Potsdam, Germany\\
$^2$Max-Planck-Institut f\"ur Radioastronomie, Bonn, Germany}
\date{} 
\end{opening}

\begin{document}

\oddpagefooter{}{}{} 
\evenpagefooter{}{}{} 
\ \bigskip

\begin{abstract}
  We give a brief review on the properties of asymptotic giant branch
  models with 
  overshoot. Then we describe new model calculations with
  overshoot. Initial masses are ranging from 1 to 3$\msun$ and
  metallicities are Z=0.02, 0.01 and 0.001. Third dredge-up occurs efficiently
  for low masses and carbon stars are formed, with some  at
  core masses 
  as low as $0.58\msun$. After the thermal pulse at which stars become C-rich the 
  luminosities are in the range of the observed C-star luminosity
  function during the whole interpulse phase and for all C-star
  models. The dredge-up evolution depends mainly on the core mass at
  the first thermal pulse and on the metallicity. The Z=0.001 models of the 2 and
  3$\msun$ sequence become 
  C-rich almost  instantaneously after the onset of the first thermal
  pulses. For 
  the 2$\msun$ case the C/O ratio initially exceeds 4. During following
  dredge-up  episodes the C/O ratio decreases.
\end{abstract}

\section{Introduction}
In recent years we have studied the influence of overshoot on the
evolution of asymptotic giant branch (AGB) stars by means of stellar modeling. Many aspects of
the evolution of AGB stars have been found to be affected by overshoot
and generally speaking the models with overshoot appear to be capable
of solving a number of problems. This is in particular related to the
ease at which they show the third dredge-up (DUP) and the modification to
the internal abundance in the region between the nuclear burning
shells (Herwig \etal1997, Bl\"ocker 1999, Mazzitelli \etal1999).

A very important application and test of the AGB models with overshoot
has been their use as a starting point for the investigation of
evolutionary scenarios of H-deficient post-AGB models (Herwig
\etal1999a). Overshooting has been able to
improve the models in two ways. First, it has been possible to
identify dredge-up as a process to cause H-deficiency of
post-AGB stars (Bl\"ocker 2000, Herwig 2000).
This opens new possibilities for the interpretation of
the observational diversity of these objects. Second, all current
evolutionary calculations of  H-deficient post-AGB stars  predict that the
abundance ratio of (He/C/O) as observable at the surface is the same as
the ratio in the intershell of the progenitor AGB star. As it appears
to date only AGB models with overshoot are sufficiently abundant in
oxygen in the intershell in order to reproduce the high observed
oxygen mass fractions of post-AGB stars (e.g.\ Werner \etal1998). These
findings support the overshoot concept for AGB and post-AGB stars in general.

\enlargethispage*{4.5cm}

\vspace{2cm}
\noindent
\begin{minipage}[h]{\textwidth}
\hrule
\medskip
\texttt{International workshop on ``The Changes in Abundances in Asymptotic Giant Branch Stars'',  
September 16th - 18th, 1999, Monteporzio Catone, Italy, to appear in {Mem. Soc.\ Astron.\ Ital.}
}
\end{minipage}
\pagebreak

While evidence for overshoot in AGB stars is growing many of the
aspects studied have been accessed only in exemplary cases (like the
$3\msun$ one with $Z=0.02$).  However, comparison with observable
properties of AGB stars as a population provide a key tool to
constrain stellar modeling. For example, the comparison of a
theoretical luminosity function (LF) of carbon stars with an observed one
allows to constrain the efficiency of DUP. In order to utilize
these constraints we  compute a new model grid covering
mass and metallicity. Previous efforts in this direction have returned
valuable insights about the order of magnitude of DUP
needed. For instance,  Marigo \etal(1999) have found that
the observed carbon star LF of the LMC can be
reproduced by models with a DUP efficiency of $\lambda=0.5$
starting at a minimum temperature at the bottom of the convective
envelope of $\log T_{\rm b}^{\rm dred} = 6.4$ which corresponds in their
analysis to a minimum core mass for DUP in the range $0.53\dots
0.58 \msun$. These studies further emphasize the need for mechanisms
to enhance DUP in stellar models of thermal pulse AGB stars because the
cited conditions cannot be met with standard models, e.g.\ those of  Forestini
\& Charbonel (1997) or Wagenhuber \& Groenewegen (1998). Clearly,
overshoot is such a mechanism.

\section{Properties of AGB stellar models with overshoot}

During our previous studies we have identified five aspects of AGB
stellar models which are modified by exponential diffusive overshoot
which we always apply to all convective boundaries.
\begin{enumerate}
\item In the region between hydrogen and
  helium burning shells (intershell region) the helium abundance is
  decreased and carbon 
  and oxygen is enhanced compared to the intershell abundances in
  models without overshoot. In particular the oxygen enhancement up to
  mass fractions of about $0.20$ is typical for models with overshoot.
  This  abundance change in the intershell is caused by a deeper
  penetration of the He-flash convection zone into the C/O core below.
  It may be referred to as ``intershell DUP'' or ``fourth DUP''.
\item This deeper penetration of the He-flash convection into the C/O
  core is closely related to the larger energy generation by
  He-burning during the thermal pulse (TP). Consequently, higher
  temperatures at the bottom of the He-flash convection zone and lower
  temperatures at the top are found, compared to models without
  overshoot.
\item These changes of structure and chemical composition in the
  intershell region substantially increase the efficiency of the
  third DUP of material from the intershell region into the
  envelope and up to the surface. The dredged-up material has a
  modified abundance distribution. More and
  different material is dredged-up than by models without overshoot.
  As a result
  models with overshoot have a different evolution of the surface
  abundances and show a stronger signature of nucleosynthesis. Due to
  efficient DUP the core mass during some more advanced TPs is
  constant or may even decrease.
\item The interpulse phase of an overshoot model sequence is
  characterized by a faster recovery of the hydrogen-burning shell
  after the TP. Consequently, the surface luminosity is recovering
  faster after the 
  TP compared to models without overshoot. 
\item At the end of the DUP a $\cdr$ and a $\nvi$ pocket at the
  envelope-core interface is formed which is presumably of importance
  for the nucleosynthesis in AGB stars.
\end{enumerate}

%

\section{Starting models and input physics}

In  our stellar evolution code (Bl\"ocker 1995, Herwig \etal1997) we now
use the most recent \textsl{OPAL} opacities 
(Iglesias \& Rogers, 1996) which have been complemented with low-temperature
tables from Alexander \& Ferguson (1994).
The initial helium mass fraction has been chosen according to
\begin{math}
Y=Y_\odot + {\Delta Y}/{\Delta Z} \cdot (Z-Z_\odot) ,
\end{math}
where $Y_\odot=0.28$ is the value adopted for the solar initial helium
abundance, $Z_\odot=0.02$ is the solar
metallicity. $\Delta Y / \Delta Z $ is the ratio of fresh helium
supplied to the interstellar  
medium by stars relative to their supply of heavy elements. Here we have
chosen a value of $\Delta Y / \Delta Z = 2.5$ (Pagel \& Portinari
1998).
The  distribution of metals is according to Grevesse \& Noels (1993) 
and Anders \& Grevesse (1989), and scaled the isotopic abundance for non-solar
metallicity. For the overshoot prescription see Bl\"ocker \etal(this volume).

We have constructed pre-main sequence starting models
from a polytropic structure
and followed the entire evolution up to the AGB.
During the main sequence evolution of models with a zero-age main
sequence mass $\mzams \leq 1.5
\msun$ we have applied a reduced overshoot efficiency of $f=0.008$
(Bressan \etal1993, Ventura \etal1998). Otherwise we used our standard value of $f=0.016$.
The evolutionary tracks  in the HRD are in  reasonable agreement with those
computed by 
Schaller  \etal(1992) with overshoot and low metallicity.
The time step during the main-sequence evolution has been limited to
1/500 of the total main-sequence evolution.

Stars with $\mzams \leq 1.7 \msun$ suffer a central He-flash at the
tip of the red giant branch (RGB). We did not calculate the central He-flash phase
explicitely. Because chemical changes induced by the flash itself can
be expected to be small, we construct zero-age horizontal branch
models by calculating equilibrium models of given mass and chemical composition
(i.e.\ mass and composition of the models at tip of RGB).

We have used a Reimers mass-loss  rate (Reimers 1975) starting at the
base of the RGB. 
For sequences with $\mzams \leq 1.7 \msun$ we choose
$\eta_{\rm R}=0.5$ and $\eta_{\rm R}=1.0$ for larger  $\mzams$. Note, that
the Reimers mass loss rate has not been designed for AGB models, and
that this choice has mainly been made for compatibility with existing
models of previous works. This approach is suitable to investigate the
main trends.

In this context we should also mention a slightly different treatment
of envelope overshoot during the actual DUP phase. Herwig
\etal(1999b) and Mowlavi (1999) found that the
DUP efficiency not much effected by variations   of the
overshoot at the bottom of the envelope convection. We
found that 
the numerical resolution can be released if a larger
overshoot parameter is applied during the actual DUP
phase. While numerical and in particular  time resolution needs to 
be very high with our original choice of the overshoot efficiency of
$f=0.016$ we have now used a ten times larger value during the period
where the actual DUP takes place. During the interpulse period
and at the bottom of the He-flash convection zone we use the original
value of $0.016$. Although a larger $f$-value can certainly not be
excluded for physical reasons we  consider it here just as a numerical method 
in order to save CPU time. Without this measure the computation
of a grid in mass and metallicity would not be possible within
reasonable time. With this increased value of $f$ we can reduce
the number of models needed to compute the DUP phase by a factor 
of ten. The trade-off for this treatment is that the DUP efficiency is larger  
by about $20\%$ compared to our original treatment, and in some cases
third DUP sets in one TP earlier than with the smaller $f$-value.
This seems contrary to the
above mentioned findings, but they do not refer to such large overshoot values.

\section{The TP-AGB models}
\label{sec:tp-agb}

We have computed 10 TP-AGB model sequences with masses ranging from
$1\msun$ to $3\msun$ and with metallicities Z=0.001, 0.01 and 0.02.
The main interest in this study was the 
occurrence of the third DUP and the subsequent formation of carbon stars
(\tab{tab:properties-preAGBI} and \ref{tab:properties-preAGBII}).
Previous studies have revealed that DUP depends on core mass,
metallicity, treatment of convection and envelope mass (Wood 1981,
Lattanzio 1986).

The  core  mass at the first TP depends on the initial stellar mass, 
on overshoot and on metallicity.
Models with main sequence core overshoot
have larger core masses than  AGB models 
without overshoot. For example the core at the first TP of our solar
$3\msun$ sequence is about $10\%$ more massive with core overshoot
than without core overshoot.
While the core masses of the Z=0.02 and 0.01 sequences are 
similar, the M2.0Z0.001 (meaning the sequence with a main-sequence mass 
of 2$\msun$ and metallicity Z=0.001) and M3.0Z0.001 sequences have much
larger core 
masses at the first TP. Due to the larger core mass the DUP is
more efficient than for models with the same initial stellar mass but
larger metallicity. The DUP of the Z=0.001
sequence is rather similar to those Z=0.02 sequences which have a
similar core mass at the first TP. 
As can be seen in \abb{fig:MHYD-AB} the M2.0Z0.001 model sequence shows a
similar DUP pattern as the solar 3$\msun$ sequence, and the M1.7Z0.02 and
the M1.5Z0.001 sequences are similar as well. Moreover we found that
the solar 4$\msun$ sequence of Bl\"ocker \etal(this volume) shows a
similar core mass evolution than our M3Z0.001 model sequence. Thus, we
conclude that the DUP efficiency of TP-AGB models with overshoot 
is most importantly ruled by the core mass  at the first TP.

\begin{figure}
    \epsfxsize=11cm 
\hspace*{0.8cm}    \epsfbox{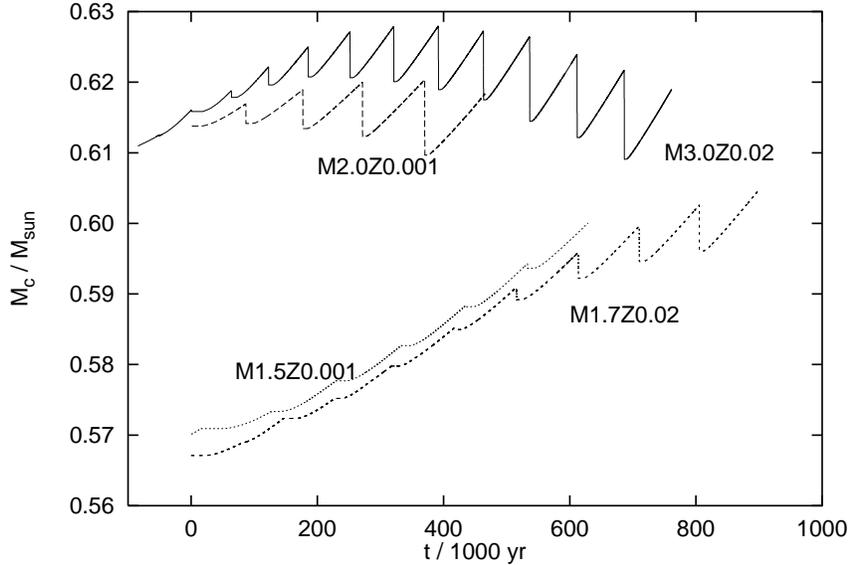}
    \caption[h]{\label{fig:MHYD-AB} Evolution of the hydrogen free core of 
    four sequences with different mass and metallicity. The core mass
    grows due to H-shell burning and is reduced by DUP.
    Models with Z=0.001 show a similar DUP pattern as models
    with solar metallicity and a roughly 20\% larger core mass.
    }
\end{figure}
Moreover DUP efficiency is directly enhanced by lower
metallicity. This is evident from the two pairs of
sequences shown in  \abb{fig:MHYD-AB}. For the Z=0.001 sequences the
DUP (and the 
interpulse period) is somewhat larger than for the Z=0.02 sequence with the
same core mass, although the
latter  has larger 
envelope masses. While the DUP difference among the Z=0.02 and
Z=0.001 case is moderate,  it is negligible among the Z=0.02 and 0.01
case. 

\begin{figure}
    \epsfxsize=11cm 
\hspace*{0.8cm}     \epsfbox{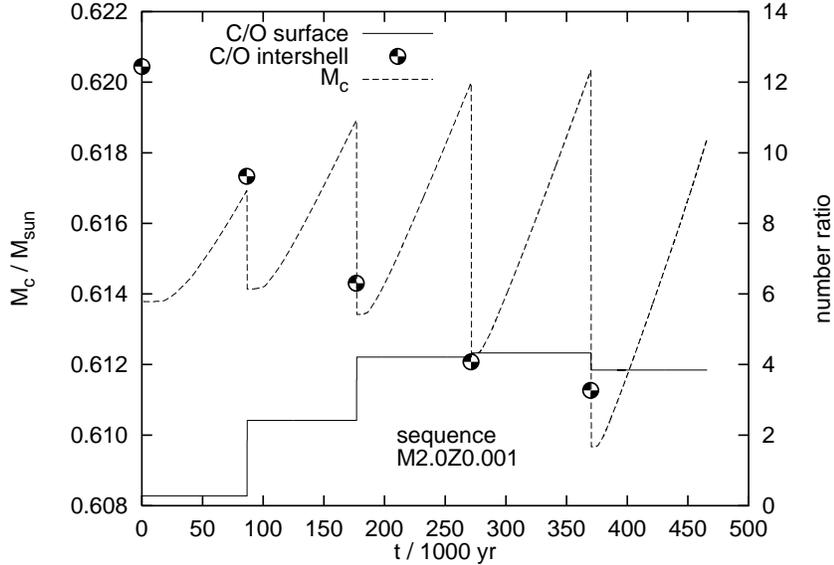}
    \caption[h]{\label{fig:C-O-ratio} Evolution of surface and intershell
    C/O ratio with time, $t=0 {\rm yr}$ at first TP. The C/O ratio at
    the surface initially 
    increases sharply after the first TP with DUP because the C/O
    ratio is very high in the intershell and the absolute mass
    fractions of C and O are small in the envelope at low Z. After a
    few TPs the C/O ratio in the intershell has decreased drastically.
    This reduction then leads to a decrease of the surface C/O ratio 
    as this intershell material is dredged-up.
    }
\end{figure}
\begin{table}[htbp]
  \begin{center}
  \caption{ \label{tab:properties-preAGBI}
    Properties of TP-AGB model sequences with overshoot. The columns
    contain sequence identifier (ZAMS mass, metallicity), core 
    and total mass at first TP,  pulse
    number, core mass and total mass of first TP with
    DUP and for first TP with number ratio ${\rm C/O \ge
    1}$.}
    \begin{tabular}{lrrrrrrrr}
      \hline
      \medskip
      sequence & 
      $M_{\rm c}^1$ &      $M_{\rm \ast}^1$ &
      $TP^{\rm min}$ & $M_{\rm c}^{\rm min}$&
            $M_{\rm \ast}^{\rm min}$ & 
      $TP^{\rm C/O \ge 1}$ & $M_{\rm c}^{\rm C/O \ge 1}$ &
      $M_{\rm \ast}^{\rm C/O \ge 1}$ \\
               &    [$\mathrm{M_\odot}$] & [$\mathrm{M_\odot}$] & &
                   [$\mathrm{M_\odot}$] & [$\mathrm{M_\odot}$] & &
                   [$\mathrm{M_\odot}$] & [$\mathrm{M_\odot}$] \\ \hline
      M1.0Z0.020 & $0.526$&$0.561$& \multicolumn{6}{c}{No DUP, end of 
      AGB      after 2nd TP.} \\ %
      M1.5Z0.020 & $0.559$&$1.240$&$(3)$ &$(0.565)$ &$(1.219)$
      &\multicolumn{3}{c}{See remark below.}  \\ 
      M1.5Z0.010 & $0.560$&$1.254$&$6$&$0.581$&$1.185$&\multicolumn{3}{c}{Last computed TP: $7$, C/O=$0.71$.}\\ 
      M1.5Z0.001 & $0.571$&$1.323$&$5
      $&$0.588$&$1.282$&\multicolumn{3}{c}{Last computed TP: $7$, C/O=$0.95$.} \\ 
      M1.7Z0.020 & $0.567$&$1.476$&$5 $&$0.585$&$1.417$&$9 $&$0.603$&$1.249$  \\ 
 M2.0Z0.020$\ast$ &$0.477$&$1.889$&$16$&$0.573$&$1.638$&$22$&$0.601$&$1.479$\\ 
      M2.0Z0.010 & $0.508$&$1.920$&$7$ &$0.534$&$1.844$&$14$&$0.578$&$1.600$\\ 
      M2.0Z0.001 & $0.614$&$1.868$&$2 $&$0.617$&$1.858$ &
      \multicolumn{3}{c}{TP(min)=TP(C/O $\ge$ 1)}\\ 
      M3.0Z0.020 & $0.616$&$2.870$&$2 $&$0.619$&$2.862$&$7$ &$0.628$ &$2.750$\\ 
      M3.0Z0.010 & $0.643$&$2.868$&$2 $&$0.646$&$2.857$&$4 $&$0.649$&$2.821$\\ 
      M3.0Z0.001 &\multicolumn{6}{c}{DUP after first TP leads to
      C/O=1.7.} &$0.813$&$0.278$ \\ 
      \hline
    \end{tabular}
  \end{center}
    \noindent
    \underline{M1.5Z0.020:} Quenching (weak thermal
    runaway without He-flash convection zone) of second TP leads to stronger
    He-flash at third TP than usual. This causes third DUP after the third
    TP. No third DUP for TP 4 and 5. Then third DUP starting continuously at TP
    6 with $M_{\rm c}^{\rm min}=0.584\msun$.
    \underline{M3.0Z0.020:} Before first TP one quenched TP.
    \underline{M2.0Z0.020$\ast$:}Computed with an older code
    version and started with too small $M_{\rm c}^1$. The new core mass 
    at the first TP  for this sequence is $0.505\msun$.
\end{table}

Eventually recurrent DUP of carbon (and oxygen) from the
intershell up to the surface leads to carbon star formation.
All but the M1.0Z0.02 sequence, listed in \tab{tab:properties-preAGBI} 
will become carbon rich. For the 1.5$\msun$ sequences C-star formation is very
likely but the computations have not been followed to the end. The
M1.0Z0.02 sequence does not experience third DUP at all. After two TPs the
AGB evolution is over because the envelope mass is lost by mass loss.

An important result are the low luminosities at the time when
carbon stars are formed. The carbon
star LF for the LMC peaks at $M_{\rm bol}=-4.9$ with
tails reaching from  $M_{\rm bol}=-3.2$ to $-5.9$ (Marigo \etal 1999,
and references there). In
\tab{tab:properties-preAGBII} we give the largest and smallest
magnitude during the interpulse phase which follows the TP of carbon
star formation. All model magnitudes fall within the observed
range. In particular, the M2.0Z0.01 sequence covers the low luminosity 
tail of the LF during the initial and shorter low luminosity phase of
the interpulse period. During the longer high luminosity phase of the
interpulse period the models have the luminosity of the LF
peak. These new
computations show that models with overshoot can produce carbon stars
of sufficiently low luminosity.

\begin{table}[htbp]
  \begin{center}
  \caption{ \label{tab:properties-preAGBII}
  Properties of TP-AGB model sequences with overshoot at that TP where
  ${\rm C/O \ge  1}$.
  The columns give sequence identifier,  bolometric magnitudes after the TP at
    which ${\rm C/O \ge 1}$ ($M_{\rm bol,min}^{\rm C/O \ge 1}$ =
    magnitude immediately after the
    TP and $M_{\rm bol,ip}^{\rm C/O \ge 1}$ =  magnitude
    towards the end of the interpulse period) DUP parameter
    $\lambda$, the age  (first TP: $t=0 {\rm yr}$)
    and the intershell abundance of helium, carbon 
  and oxygen at this pulse.}
  \medskip
    \begin{tabular}{lrrrrr}
      \hline
      \medskip
      sequence & 
      $M_{\rm bol,min}^{\rm C/O \ge 1}$ &
      $M_{\rm bol,ip}^{\rm C/O \ge 1}$ &
      $\lambda^{\rm C/O \ge 1}$ &
      $t^{\rm C/O \ge 1}$ &
      $(He/C/O)_{\rm is}^{\rm C/O \ge 1}$ 
      
 \\ 
               & [mag]        & [mag] &
      [ratio] & [yr] & [mass fractions] \\ \hline
      M1.7Z0.020 &$-4.31$&$-5.14$&$0.81$&$4.16{\rm E}5$ &$(0.32/0.45/0.20)$  \\ 
M2.0Z0.020$\ast$ &$-4.38$&$-5.16$&$0.65$&$4.09{\rm E}6$&$(0.33/0.43/0.20)$\\
      M2.0Z0.010 &$-4.06$&$-4.84$&$0.54$&$2.30{\rm E}6$&$(0.25/0.42/0.32)$\\
      M2.0Z0.001 &$-4.08$&$-4.77$&$0.87$&$8.64{\rm E}4$&$(0.44/0.49/0.07)$\\
      M3.0Z0.020 &$-4.63$&$-5.28$&$1.13$&$3.92{\rm E}5$ &$(0.29/0.44/0.23)$\\ 
      M3.0Z0.010 &$-4.65$&$-5.25$&$1.36$&$1.75{\rm E}5$&$(0.29/0.52/0.17)$\\
      M3.0Z0.001 & $-5.39$&$-5.62$ &$1.90$ &$0$&$(0.59/0.37/0.04)$ \\ 
      \hline
    \end{tabular}
  \end{center}
\end{table}
The formation of carbon stars can proceed over many TPs with small
increases of the C/O ratio at each TP. For example,  the M2.0Z0.010
sequence needs 7 TPs from the start of DUP until the model is carbon
rich. The other extreme are the more massive and very low metallicity
model sequences which can become carbon rich at the first or second TP 
where DUP is instantly present. These models allow to study the
variation of the surface C/O ratio as a function of pulse number which 
is of relevance for less massive and more metal rich models as well.
This variation is displayed in \abb{fig:C-O-ratio} for the M2.0Z0.001
sequence. 
The C/O ratio in the intershell is very high during the first TP and
decreases continuously over the following TP. This high intershell C/O 
ratio together with the lower absolute mass fractions of C and O in
the envelope for Z=0.001 leads to an immediate transformation into a
carbon star during the first DUP episode after the second TP.
Following DUP episodes are even more
efficient with $\lambda>1$ and the C/O number ratio quickly reaches
values as high as $4$. However, at following TPs
the C/O ratio is not increased by DUP but decreased.
This is because the C/O ratio in the
intershell is decreasing at each TP during the first TP-AGB phase
when more oxygen is mixed from the C/O core into the intershell
region by intershell DUP. At the fourth TP of  sequence M2.0Z0.001
(\abb{fig:C-O-ratio}) the surface C/O ratio has reached the intershell 
C/O ratio. During the following TP the intershell C/O ratio will
decrease further and then \emph{DUP will decrease the surface C/O
ratio}. For less massive or more metal rich stars the formation of
C-stars is not fast enough in order to exhibit a surface C/O ratio
maximum. However, it is clear that the surface C/O ratio in these
stars will also asymptotically approach the intershell
ratio, which is believed to show up at the surface of H-deficient
post-AGB stars. Therefore, overshoot AGB models cannot produce too large C/O
ratios, even if the mass loss is very small, as long as the intershell 
C/O ratio is small enough. Note, that the largest observed C/O
ratio in stars  or PNe (Zuckermann \& Aller 1986) is of the same order 
as the typical asymptotic C/O intershell ratio.  

\section{Conclusion}
We have studied the TP-AGB evolution of low mass stars as a function
of metallicity. We find that C-star production at low core masses is
possible with overshoot. Typically, carbon star models of low mass 
stars have core masses around 0.6$\msun$. The C/O ratio
asymptotically reaches the value of the intershell which is around 3
in models with overshoot. This may be a natural explanation why the
largest observed C/O ratio in stars or PNe is actually at about this
value. These models should help to evaluate the appropriateness of the 
overshoot concept for TP-AGB stars.

\acknowledgements This work has been supported by the \emph{Deut\-sche
  For\-schungs\-ge\-mein\-schaft, DFG\/} (La\,587/16).

\end{document}